# Incipient spin-dipole coupling in a 1D helical-chain metal-organic hybrid

*Tathamay Basu\*, Clarisse Bloyet, Jean-Michel Rueff, Vincent Caignaert, Alain Pautrat and Bernard Raveau\**

Normandie Univ, ENSICAEN, UNICAEN, CNRS, CRISMAT, 6 Bd du Maréchal Juin, 14050 Caen Cedex, France.

*Guillaume Rogez*

IPCMS, UMR Unistra-CNRS 7504, 23 rue du Loess, BP 43, 67034, Strasbourg Cedex 2, France.

*Paul-Alain Jaffrès*

CEMCA UMR CNRS 6521, Université de Brest, IBSAM, 6 Avenue Victor Le Gorgeu, 29238 Brest, France.

\*Email address: tathamaybasu@gmail.com; bernard.raveau@ensicaen.fr

## *Abstract*

Low dimensional magnetic systems (such as spin-chain) are extensively studied due to their exotic magnetic properties. Here, we would like to address that such systems should also be interesting in the field of dielectric, ferroelectricity and magnetodielectric coupling. As a prototype example, we have investigated a one-dimensional (1D) helical-chain metal-organic hybrid system with a chiral structure which shows a broad hump in magnetic susceptibility around 55 K ($T_{max}$). The complex dielectric constant exactly traces this feature, which suggests intrinsic magnetodielectric coupling in this chiral system. The dipolar ordering at $T_{max}$ occurs due to lattice-distortion which helps to minimize the magnetic energy accompanied by 1D-magnetic ordering or vice-versa. This experimental demonstration initiates a step to design and investigate hybrid organic-inorganic magnetic systems consisting of chiral structure towards ferroelectricity and magnetodielectric coupling.

Low dimensional magnetic systems, such as, spin-chain, spin-ladder, single molecule magnets, have received considerable interest in the past years because they exhibit many fascinating phenomena due to the presence of exotic magnetism, such as spin-gap, spin-Peierls instability and slow relaxation of the magnetization.[1–5] Tremendous research has been performed in both organic and inorganic systems from the magnetism point of view, however, those materials could be equally important in the field of dielectric, ferroelectricity, piezoelectricity, multiferroicity and magnetodielectric coupling. Ferroelectricity was demonstrated in a charge-transfer salt tetrathiafulvalene-p-bromanil, a probable one-dimensional organic quantum magnet, where it is evident that spin-Peierls instability plays an important role to generate ferroelectricity.[6] An electric dipole driven magnetism was also documented in a charge-transfer organic salt, where long range magnetic and ferroelectric ordering is stabilized *via* loss of magnetic frustration.[7] Recently, it has been speculated that low dimensional magnetism could also favor magnetodielectric coupling, which is one of the current hot topics in the field of solid state physics and material science, through investigation of some spin-chain oxide systems.[8–11] The magnetic instability often helps lattice distortion to minimize the total ground state energy, so that multiferroicity/magnetodielectric coupling arises in those systems. Eventually, among those few available reports in organic/inorganic systems, a direct one-to-one correlation between magnetism and dielectric in a low-dimensional (spin-chain) system was practically demonstrated only in the geometrically frustrated spin-chain oxide $Ca_3Co_2O_6$, by documenting a clear broad hump in temperature dependent magnetic susceptibility and dielectric constant.[9,10]

Apart from pure organic/inorganic materials, another recent promising approach is to design multiferroic/magnetodielectric hybrid organic-inorganic frameworks (HOIF), where one can flexibly play with different ligands to modify the network to tune magnetic and electric properties simultaneously.[12–17] However, very little research exists from the viewpoint of multiferroicity or/and magnetodielectric coupling in hybrid organic-inorganic frameworks and rarely in magnetically low-dimensional HOIF systems. Bearing in mind that the basic criterion for the appearance of ferroelectricity deals with the breaking of spatial inversion symmetry, HOIF materials involving helical chains have a great potential. One can indeed design a system with/without inversion center of the helix to obtain a centrosymmetric/ non-centrosymmetric compound, though very little attention were paid.[18–20] However, existence of local dipoles does not imply that a macroscopic polarization can be always observed, especially when static disorder is important. Appearance of ferroelectricity is reported in a designed chiral metal-organic-framework, $(CH_3NH_3)_{12}\{Cu^{II}_{24}[(S,S)\text{-hismox}]_{12}(OH_2)_3\}\cdot 178\ H_2O$.[21] The advantage of HOIF's with respect to pure inorganic oxides, is that one can design a chiral non-centrosymmetric compound. To date, there is no report on multiferroicity/magnetoelectric coupling on such a chiral system, though the chiral crystal structure containing a magnetic ion could be a good candidate for such investigations.

Here, we have investigated a 1D helical chain HOIF system, $Cu_6(H_2O)_7(m\text{-}PO_3\text{-}C_6H_4\text{-}COO)_4$,[22] through detailed dielectric measurements. We have specifically chosen this HOIF system to investigate dielectric properties, because, i) it is a chiral material which is non-centrosymmetric, and ii) it is a magnetically low-dimensional system built up of helical chains. This compound crystallizes in the hexagonal non-centrosymmetric space group $P6_522$, and consists of two kinds of $Cu^{2+}$-ions, forming bipyramidal $Cu_2O_9$ units and single $CuO_5$ pyramids.[22] Two $Cu_2O_9$ units are connected by one $CuO_5$ pyramid which leads to a helix of corner sharing $CuO_5$ pyramids running along the $c$ axis (as shown in figure 1). The Cu-helical chains are separated by phosphono-carboxylate groups, which makes the structure one dimensional.[22] Moreover, it was previously shown[22] that the material was likely isolated as a conglomerate[20] indicating that in each crystal the helical chains are always turning either anticlockwise or clockwise.[22] Accordingly each crystal possesses a non-centrosymmetric structure.[22] This compound exhibits a broad hump in temperature dependent susceptibility around 55 K, which is characteristic of 1D magnetic behavior.[22]

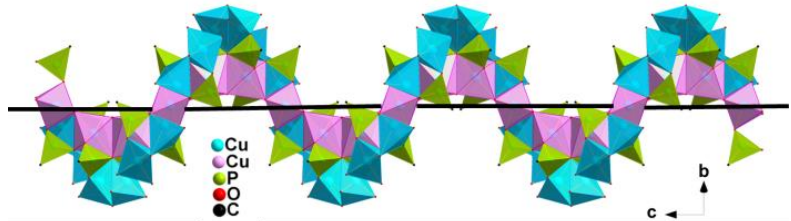

**Figure 1:  A chiral helix view of copper chains of $Cu_6(H_2O)_7(m\text{-}PO_3\text{-}C_6H_4\text{-}COO)_4$ along *a*. $CuO_5$ pyramids (blue colored) share corners with bi-pyramidal $Cu_2O_9$ units (pink colored). The $PO_3C$ phosphonate tetrahedra are represented in green. The benzene rings that are functionalized with a phosphonic acid and a carboxylic acid groups are omitted for clarity.[22]**

The sample has been synthesized by hydrothermal method, as described by Rueff et al.[22] The powder X-ray diffraction (PXRD) has been performed to check the phase purity and structure (see figure 2), which completely agrees with the earlier report.[22] We have also performed PXRD before and after magnetic measurement to cross-check the stability of the sample, PXRD pattern was reproducible confirming the phase stability of the sample. The magnetic measurement has been performed using a commercial Superconducting Quantum Interference Device (SQUID) magnetometer (Quantum Design). The dielectric measurement has been carried out as a function of temperature (*T*) and magnetic field (*H*) using a impedance analyzer (LCR meter E4886A, Agilent Technologies) which is integrated into a commercial Physical Properties Measurement System (PPMS, Quantum Design). The parallel plate capacitor was made by painting silver paste on both sides of a pellet (5mm diameter, 1mm thickness). The pellet was made by crushing small single crystals applying low pressure without sintering to avoid any deterioration/ degradation of the metal-organic sample. Therefore, this is a soft pellet, not a strong dense sintered pellet. The experiments have been repeated for different batches of sample to ensure the intrinsic dielectric behavior of this compound.

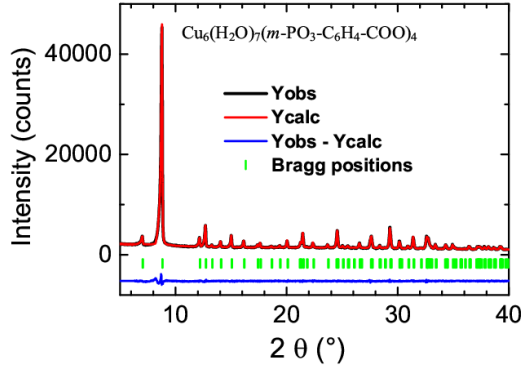

**Figure 2: Powder X-ray diffraction (PXRD) at room temperature for the compound $Cu_6(H_2O)_7(m\text{-}PO_3\text{-}C_6H_4\text{-}COO)_4$.**

The dc-magnetic susceptibility has been already reported by Rueff et al.,[22] however, we have reproduced the same to compare with dielectric results. The temperature dependent dc susceptibility (here, $\chi=M/H$) is shown in figure 3a for the magnetic field of 5 kOe, and completely agrees with earlier report.[22] The Curie-Weiss temperature (-27 K) obtained from Curie-Weiss fitting in paramagnetic region (not shown here, see Ref. [22]) indicates the presence of antiferromagnetic interactions in this S= ½ system. The paramagnetic Curie-Weiss behavior deviates below 150 K and shows a broad hump around 30-70 K with a maximum ($T_{max}$) ~ 55 K,.[22] Such a broad feature is typical of short-range antiferromagnetic interactions in a low-dimensional magnetic system.[22] In addition, we have measured ac susceptibility. Its imaginary part is zero throughout the temperature range (not shown here) what is consistent with the absence of long range ordering of spins or spin-glass behavior.

The temperature dependent real ($\varepsilon'$) and imaginary ($\varepsilon''$) parts of complex dielectric constant for 21 kHz are shown in figure 3b and 3c respectively. The low value of loss part ($\tan\delta \ll 1$, as shown in the inset of figure 3c) confirms the highly insulating nature of this sample and excludes any extrinsic effect like leakage current, which is a necessary condition to investigate dielectric properties. The real part of the dielectric constant $\varepsilon'(T)$ decreases with decreasing temperature from high temperature yielding a clear peak around $T_{max}$ (55K), tracing the maximum in magnetic susceptibility. The dielectric feature is reproduced for other frequencies as well confirming the intrinsic nature of this effect (see inset of figure 3b). The compound is already characterized as a non-centrosymmetric system, suggesting that dipoles order at $T_{max}$, which yields such a peak. The height of the peak is weak, which could be due to the soft nature of the pellet of the metal-organic sample. This feature was reproduced for different batches of the sample. The imaginary part of the dielectric constant $\varepsilon''(T)$ shows a broad peak from 30-70 K, exactly mimicking the broad hump in $\chi(T)$. These results manifest that both the features in dielectric and magnetization at $T_{max}$ should have the same origin.

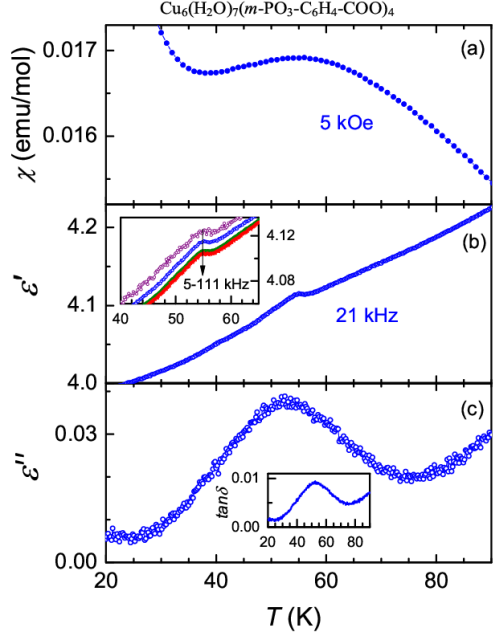

**Figure 3: (a)** Dc susceptibility as a function of temperature for 5 kOe magnetic field. **(b)** Real and **(c)** imaginary part of dielectric constant as a function of temperature from 10-90 K for a fixed frequency of 21 kHz for the compound $Cu_6(H_2O)_7(m\text{-}PO_3\text{-}C_6H_4\text{-}COO)_4$; inset of (b) shows the $\varepsilon'(T)$ for different frequencies; inset of (c) shows loss part ($tan\delta$) as a function of $T$ for 21 kHz.

Here, 1D-magnetic ordering occurs at 55 K *via* distortion of the lattice (change in magnetic-path length along the helical-chain) which helps to increase the magnetic exchange interaction along the spin-chain. Also, this lattice distortion helps to order the dipoles. Thus, the overall ground state energy of this system is minimized as a result of lattice distortion. Therefore, the magnetic ordering is accompanied by dipolar ordering or vice-versa. Such a role of 1D-magnetic correlation with spin-chain to create significant ferroelectricity has been theoretically predicted for the multiferroic Haldane-chain compound $Er_2BaNiO_5$ (see Ref. [8]). We also endorse such a possibility. We did not observe any significant effect of external magnetic field on the dielectric constant within the resolution limit of instrument, the excess dielectric constant with variation of magnetic field is <0.01%. Bearing in mind that the application of an external magnetic field has no effect on the magnetic behavior of this compound, due to its 1D magnetic nature, we can expect that the effect of magnetic field upon the dielectric constant is also negligible. A negligible change of magneto-elastic coupling with varying magnetic field will cause a negligible change in dielectric constant. Such an incipient spin-dipole correlation in a 1D magnetic system has been rarely demonstrated. It was experimentally observed in the oxide compound, $Ca_3Co_2O_6$,[9,10] which is a geometrically frustrated spin-chain oxide and exhibits long-range magnetic ordering at lower temperature.[9,10] It was also proposed that ferroelectricity can be induced by low-dimensional instability in the charge-transfer organic salt tetrathiafulvalene-p-bromanil.[6] However, in this organic salt, one dimensional array of spins arises from alternating arrangement of unpaired donor (D)-acceptor (A) ionic ligands. A paramagnetic to non-magnetic transition is observed due to dimerization of D-A,[23] which induces ferroelectricity. Therefore, this organic system is not characterized as a typical 1D magnet. Nevertheless, these results suggest that ferroelectricity should be generated in favorable circumstances in a 1D magnetic system. We could not confirm the ferroelectricity in our system, as pyroelectric or polarization measurements are impeded due to the soft polycrystalline nature of this metal-organic sample (not a strong dense pellet due to absence of sintering as discussed earlier in experimental part). A large single crystal would be needed to perform such experiment to confirm ferroelectricity.

This result reveals that magnetically low-dimensional systems should be good candidates to explore ferroelectricity and cross-coupling between spin and dipoles. The chiral structure is responsible for spatial inversion symmetry breaking which creates local dipoles, and orders at further low temperature (55 K) accompanied by magnetic ordering. Therefore, the chiral structure also plays an important role for dipolar ordering and spin-dipole coupling. Further microscopic and theoretical investigations are highly warranted for deeper understanding of spin-dipole correlation in this helical-chain magnet.[24–27]

In summary, we have demonstrated that the 1D helical-chain magnet $Cu_6(H_2O)_7(m-PO_3-C_6H_4-COO)_4$ exhibits a clear dielectric anomaly at the onset of magnetically low-dimensional ordering. Such a phenomenon, rarely reported for 1D systems and observed herein for the first time for metal-organic hybrid material, can be attributed to the existence of dipole ordering originating from magnetodielectric coupling. The latter may appear via lattice-distortion by minimizing the overall ground state energy for a system exhibiting 1D-magnetic ordering. This observation of incipient spin-dipole coupling in a hybrid metal-organic 1D helical chain magnet paves the way for the investigation of other hybrid organic-inorganic materials for the generation of 1D magnetoelectric system and ferroelectricity. Also, our result opens up the possibility to search for magnetodielectric coupling and multiferroicity in metal-organic magnets with a chiral structure.

**Conflicts of interest:**
There are no conflicts of interest to declare.

**Acknowledgements:**
We thank the Agence Nationale de la Recherche (contract N°ANR-14-CE07–0004–01 (HYMN)) for financial support. The authors also express their grateful acknowledgment for technical support to Fabien Veillon from the CRISMAT laboratory.